\documentclass[aps,prb,twocolumn,superscriptaddress,10pt]{revtex4-1}

\usepackage{graphicx}
\usepackage{amsmath}
\usepackage{amssymb}
\usepackage{amsfonts}
\usepackage{bm}
\usepackage{psfrag}
\usepackage{pstricks}
\usepackage{multirow}

\usepackage{color}

\begin{document}
\title{Polar InGaN/GaN quantum wells: Revisiting the impact of carrier localization on the green gap problem}

\author{Daniel S.~P. Tanner}
\affiliation{Photonics Theory Group, Tyndall National Institute,
University College Cork, Cork, T12 R5CP, Ireland}
\author{Philip Dawson}
\affiliation{School of Physics and Astronomy, University of
Manchester, Manchester, M13 9PL, United Kingdom}
\author{Menno J. Kappers}
\affiliation{Department of Materials Science and Metallurgy, 27
Charles Babbage Road, University of Cambridge, Cambridge, CB3 0FS,
United Kingdom}
\author{Rachel A. Oliver}
\affiliation{Department of Materials Science and Metallurgy, 27
Charles Babbage Road, University of Cambridge, Cambridge, CB3 0FS,
United Kingdom}
\author{Stefan Schulz}
\affiliation{Photonics Theory Group, Tyndall National Institute,
University College Cork, Cork, T12 R5CP, Ireland} 

\begin{abstract}
We present a detailed theoretical analysis of the electronic and
optical properties of $c$-plane InGaN/GaN quantum well structures
with In contents ranging from 5\% to 25\%. Special attention is paid
to the relevance of alloy induced carrier localization effects to
the green gap problem. Studying the localization length and
electron-hole overlaps at low and elevated temperatures, we find
alloy-induced localization effects are crucial for the accurate
description of InGaN quantum wells across the range of In content
studied. However, our calculations show very little change in the
localization effects when moving from the blue to the green spectral
regime; i.e. when the internal quantum efficiency and wall plug
efficiencies reduce sharply, for instance, the in-plane carrier
separation due to alloy induced localization effects change weakly.
We conclude that other effects, such as increased defect densities,
are more likely to be the main reason for the green gap problem.
This conclusion is further supported by our finding that the
electron localization length is large, when compared to that of the
holes, and changes little in the In composition range of interest
for the green gap problem. Thus electrons may become increasingly
susceptible to an increased (point) defect density in green emitters
and as a consequence the nonradiative recombination rate may
increase.
\end{abstract}

\date{\today}

\maketitle

\section{Introduction}

\begin{figure}
\includegraphics{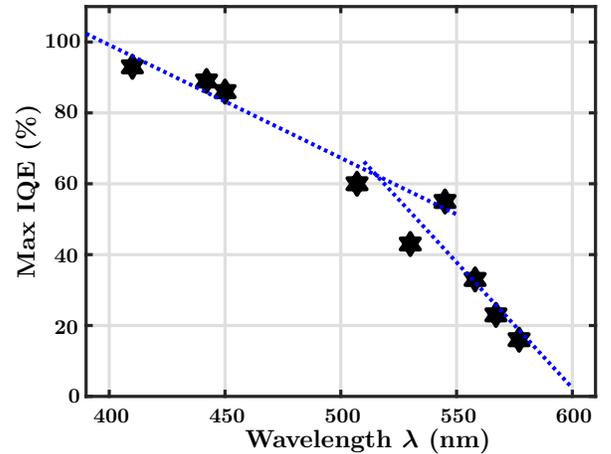}
\caption{Literature values for maximum internal quantum efficiency
(IQE) values as a function of the emission wavelength of $c$-plane
InGaN/GaN systems.~\cite{Karp2016}} \label{fig:MaxIQE_intro}
\end{figure}

Wurtzite InGaN/GaN quantum well (QW) systems, grown along the polar
$c$-axis, have allowed the realization of optoelectronic devices
that emit bright light over a wide spectral range, including
emission at previously unattainable short wavelengths, with
unprecedented efficiency.~\cite{PoBo1997,AkAm06,NaMu93} However, the
internal quantum efficiency (IQE) drops rapidly for emission
wavelengths beyond the blue spectral
range,~\cite{MuYa99,Humphreys_MRS} as shown in
Fig.~\ref{fig:MaxIQE_intro}. This effect is known as the ``green
gap'' problem.~\cite{Humphreys_MRS,KrSc07} Given the dramatic
reductions in energy use, and the associated reduction in carbon
emissions which would result from the development of all-LED white
light sources, utilizing red, green, and blue
LEDs,~\cite{Humphreys_MRS} the understanding and circumvention of
the green gap problem represents an urgent scientific and ecological
objective. However, the origin of this phenomenon in InGaN-based
emitters is still a matter of some debate. Focusing on $c$-plane
InGaN/GaN QWs, the green gap problem has been discussed on the basis
of mainly three factors and their combined effect on the
radiative/nonradiative recombination processes in these
heterostructures.

The first of these factors is the decreasing material quality of
InGaN/GaN QW systems with increasing In
content.~\cite{Crawford,LaHo13,ArHe2012,MaDa2014,UeIs2014,HaKa2015}
To grow InGaN QWs emitting in the green spectral region, low growth
temperatures are typically required to suppress In atom desorption
during growth and thus to achieve higher In
contents.~\cite{HaKa2015} The results by Hammersley \emph{et
al.}~\cite{HaKa2015} indicated that these low growth temperatures
lead to increased (point) defect densities. Additionally, with
increasing In content in the well, the lattice mismatch between the
InGaN QW and the GaN barrier increases, which encourages stress
relaxation through defect formation.~\cite{Crawford,LaHo13,MaDa2014}
In general, defects and structural imperfections can serve to reduce
the IQE by acting as nonradiative recombination
centers.~\cite{LaHo13,LaKr11} Some of these issues may be reduced
using innovative growth techniques.~\cite{ArBr15,ZhLu17,TiLi17}

The second factor contributing to the green gap problem is a
property intrinsic to wurtzite $c$-plane InGaN/GaN QWs and is also
related to the increase in lattice-mismatch induced strain with
increasing In content. The increase in strain leads to an increase
in the piezoelectric polarization vector field inside the InGaN
QW.~\cite{AmMa2002,PaSc2017} The resultant intrinsic electrostatic
built-in field leads to a separation of electron and hole wave
functions along the growth direction. This effect is also known as
the quantum confined Stark effect (QCSE).~\cite{AnORe2001} Thus, the
higher the In content the larger the built-in field and the greater
the reduction in the radiative recombination rate due to the
separation of charge carriers.

Taken together, factor one (increased defect densities and
nonradiative recombination rates), and factor two (reduction in
radiative recombination rate) will have a negative impact on the
radiative optical properties of InGaN-based light emitters. As
discussed above, these limitations will be worsened for higher In
contents and thus emission in the green spectral region.

Very recently, a third factor contributing to the green gap problem
has received significant attention: carrier
localization~\cite{YaSh2014,AuPe2016,JoTe2017,Karp2017,Karp2018,NiKa16}
due to (random) alloy fluctuations as well as structural
inhomogeneities. Different groups have concluded that the alloy
related carrier localization effects lead to an in-plane separation
of electrons and holes in addition to the built-in field induced
effect, described above.~\cite{AuPe2016,Karp2017,Karp2018,WaGo2011}

Despite the large quantity of recent studies on this topic, there
remains a significant degree of uncertainty as to which contribution
is the dominant factor for the green gap problem, and accordingly
how it may be best circumvented. Many works impute the green gap
primarily to an increased defect
density.~\cite{LaKr11,LaHo13,DaYo2019} Other studies focus on
carrier localization effects as the main source of this
effect,~\cite{NiKa16,MaKa2017,Karp2018} and assume that
non-radiative recombination effects (such as defect-related
Schockley-Read-Hall recombination) are independent of the In
content. These recent theoretical studies, whereby the green gap is
attributed to material intrinsic carrier localization result in a
very pessimistic outlook in which the green gap cannot be closed via
technological optimization of InGaN QWs.~\cite{MaKa2017,Karp2018}
This emphasis on localization as a negative feature of InGaN QWs is
an interesting development, given that localization is widely
regarded as the mechanism by which InGaN-based light emitting
devices, with their high defect densities, exhibit high efficiencies
at all.~\cite{ChUe2006,DaSc2016}

Given this ambiguity and disagreement in the literature, and the
distinct approaches that each of these hypotheses suggest for
further development of InGaN technologies, we revisit in this work
the contentious issue of the origin of the green gap. Our results
show that the localization characteristics in InGaN QWs do not
change strongly from blue to green emitters, but rather change very
little in this wavelength range. This lack of variation in
localization effects has not been considered or discussed in works
attributing the green gap problem to carrier localization; only
absolute numbers are discussed rather than focussing on changes in
the carrier localization effects. Given that blue LEDs are very
efficient, and that alloy induced carrier localization effects do
not change dramatically between blue and green emitters, only slight
changes to the \emph{radiative} recombination rate would be
expected. This casts doubt on the assumption that the nonradiative
recombination due to defects is constant and In content independent;
if it were, the sharp drop in max IQE (see
Fig.~\ref{fig:MaxIQE_intro}) and/or wall plug efficiency (WPE), as
we will discuss below, would be difficult to explain. So contrary to
the pessimistic conclusion that due to alloy induced carrier
localization effects the green gap problem cannot be solved,
innovative growth techniques may allow for its closure. For example,
Haller~\emph{et al.}~\cite{HaCa2017} have recently suggested that by
growing an InGaN underlayer prior to the QW layers, it is possible
to getter nonradiative point defects to the underlayer, thereby
increasing efficiency.

In this work we come to our conclusion by using an atomistic
multi-band tight-binding (TB) model that we first rigourously
benchmark against experimental data for InGaN/GaN QWs emitting over
a wide range of emission wavelengths. We study the photoluminescence
(PL) peak energies, widths, and carrier localization lengths.
Overall very good agreement between theory and experiment is found.
This highlights that carrier localization effects are accurately
described by our model, thus providing a solid foundation for
analyzing and predicting the importance of carrier localization for
the green gap problem.

The manuscript is organized as follows. In Sec.~\ref{sec:Theory} we
introduce the theoretical framework. In Sec.~\ref{sec:QW_structures}
structural properties of QW structures for the different In contents
will be discussed along with a description how these are treated in
the modeling. The results of our calculations are presented in
Sec.~\ref{sec:Results}. A theory experiment comparison is given in
Sec.~\ref{sec:TheoVsExp}. The analysis of the carrier localization
length is presented in Sec.~\ref{sec:CarrierLocLength}, followed by
investigating the electron-hole wave function overlap as a function
of In content in Sec.~\ref{sec:WFoverlaps}. Our work is summarized
in Sec.~\ref{sec:Conclusion}. Further details of the calculations
and further supporting results are given in an Appendix.

\begin{table*}[t!]
\caption{Experimental and theoretical data on structural (Indium
content, well width $L_w$) and optical properties (photoluminescence
(PL) peak energy, full width at half maximum (FWHM)) of single
InGaN/GaN QWs with varying In content. The experimental data
(Experiment) have been adapted from Ref.~\onlinecite{GrSo2005} (see
also Ref.~\onlinecite{Wats2011}). The results from calculations in
the framework of a modified 3D single-band effective mass
approximation are denoted by EMA.~\cite{WaGo2011,Wats2011} The
outcome of the present study, using atomistic tight-binding theory,
are labeled by TBM.} \label{Tab:QW_results_EMA_EXP}
\begin{tabular}{|c||ccccc||ccccc||cccc|}\hline
\textbf{Quantity}& \multicolumn{5}{|c||}{\textbf{Experiment}} &
\multicolumn{5}{c||}{\textbf{Theory EMA}} &
\multicolumn{4}{c|}{\textbf{Theory TBM}}
\\\hline
Indium content (\%) & 5$\pm$3 & 12$\pm$3 & 15$\pm$3 & 19$\pm$2 &
25$\pm$2 & 5 & 12 & 15 & 19 & 25 & 5 & 10 & 15 & 25
\\\hline
Well width $L_w$ (nm)     & 2.5$\pm$0.3 & 2.7$\pm$0.3 & 2.9$\pm$0.3
& 3.2$\pm$0.2 & 3.3$\pm$0.2 & 2.5 & 2.7 & 2.9 & 3.2 & 3.3 & 2.85 &
2.9 & 3 & 3.5
\\\hline
PL peak energy (eV) & 3.32 & 2.99 & 2.71 & 2.36 & 2.16 & 3.27 & 2.88
& 2.69 & 2.49 & 2.14 & 3.23 & 2.96 & 2.62 & 1.99\\\hline FWHM (meV)
& 23 & 62 & 76 & 56 & 61 & 34 & 50 & 58 & 69 & 75 & 35 & 59 & 84 &
93\\\hline
\end{tabular}
\end{table*}

\section{Theory and quantum well system}
\label{sec:Theory}

In this section we briefly review the theoretical framework used and
the QW model systems studied. We start in Sec.~\ref{sec:Theory} with
the theoretical framework and briefly introduce its main
ingredients. In Sec.~\ref{sec:QW_structures} the supercell employed
in our calculations is described.

\subsection{Theoretical framework}
\label{sec:Theory}

Our theoretical framework is based on an atomistic nearest neighbor
$sp^3$ TB model. The matrix elements of the TB Hamiltonian
describing the binary materials are treated as free parameters,
which have been obtained by fitting the TB band structures to hybrid
functional density functional theory data.~\cite{CaSc2013local}
While for In and Ga atoms the nearest neighbor environment always
consists of N atoms, N atoms can have variable numbers of In and Ga
atoms as nearest neighbors. To determine the N-onsite energies in an
alloy, weighted averages of the InN and GaN binary values have been
used, which is a widely used approximation in TB models for
alloys.~\cite{OReLi2002,LiPo92,BoKh2007}

To achieve an atomistic description of the strain field in a
strained InGaN QW, we utilize a valence force field (VFF)
model.~\cite{CaSc2013local} By minimizing the VFF elastic energy of
the simulation supercell, the relaxed atomic positions of the system
are determined. These then serve as input for the TB model, which
accounts for the effect of strain on the electronic structure by an
on-site correction to the matrix elements, the method of which is
presented in Ref.~\onlinecite{CaSc2013local}.

In addition to (local) strain effects, (local) built-in polarization
fields also affect the electronic and ultimately the optical
properties of InGaN/GaN QWs.~\cite{WiAn2005,ScCa2015,ScTa2015} In
these systems in particular the strain induced piezoelectric
contribution plays a major
role.~\cite{WiAn2005,PaSc2017,AnORe2001,PaWa2017} Macroscopic, as
well as local polarization effects are included in the calculation
via the local polarization theory described in
Ref.~\onlinecite{CaSc2013local}.

By diagonalizing the constructed TB Hamiltonian matrix, the
corresponding single-particle states are obtained. In previous
work,~\cite{ScCa2015,TaMcM2018,DaSc2016} we have shown that for
$c$-plane InGaN/GaN QWs at low temperatures, especially for well
thicknesses exceeding 2.5 nm, Coulomb effects mainly lead to
energetic shifts of the peak positions in the optical spectra but do
not alter the localization characteristics significantly. Given that
we are interested in carrier localization effects  and
characteristics, neglecting Coulomb (excitonic) effects is a
reasonable starting point for our analysis. To study the optical
properties of $c$-plane InGaN/GaN QWs with varying In content at low
temperatures, we use single-particle states and energies in
conjunction with Fermi's golden rule to obtain PL emission
spectra.~\cite{ScSc2006,ScCa2015} For low temperature and low
carrier density studies we calculate the dipole matrix elements for
electron and hole ground states for each microscopic alloy
configuration. Even though a large number of microscopically
different configurations have been used (see below), in Fermi's
Golden rule the $\delta$-function, describing the energy
conservation, has been replaced by a Lorentzian function with a
standard deviation of 7 meV. For the composition range studied and
for the number of considered microscopic configurations, this
setting results generally in smooth low temperature PL spectra, with
the broadening of the $\delta$-function having a minimal effect on
the width of the calculated spectra.

Finally, the supercell that contains the InGaN/GaN QW system and on
which the calculations will be performed, has to be defined. This
will be the topic of the next section.

\subsection{Quantum well structures and simulation supercell}
\label{sec:QW_structures}

To study the importance of alloy-induced carrier localization to the
green gap problem, we have chosen to model $c$-plane InGaN/GaN QWs
with In contents of 5\%, 10\%, 15\%, and 25\%, covering the blue to
green spectral range. We simulate these QW systems using supercells
with periodic boundary conditions that contain 81,920 atoms,
corresponding to a system size of approximately $10\times9\times10$
nm$^3$. The details of the structures used in our theoretical study
are labeled as ``TBM'' in Table~\ref{Tab:QW_results_EMA_EXP}.

To compare and benchmark our results against literature, we make use
of Refs.~\onlinecite{WaGo2011},~\onlinecite{GrSo2005}
and~\onlinecite{Wats2011}, where the optical properties of a series
of $c$-plane InGaN/GaN QWs have been studied in theoretical and
experimental works. The theoretical results of
Ref.~\onlinecite{Wats2011} were obtained in the framework of a
modified 3D single-band effective mass approach, further details of
which may be found in Ref.~\onlinecite{Wats2011}. The details of the
studied QWs are also presented in
Table~\ref{Tab:QW_results_EMA_EXP}; experimental data from Graham
\emph{et al.}~\cite{GrSo2005} are entitled ``Experiment'', while the
results of the modified 3D single band effective mass approximation
are labeled ``EMA''. Since our focus is on general trends with
increasing In content, we do not aim for a direct one-to-one
comparison with the structures reported in
Ref.~\onlinecite{GrSo2005}. Such a one-to-one theory-experiment
comparison would in general be difficult, given that the In
fractions and well widths reported in Ref.~\onlinecite{GrSo2005}
have been determined by electron energy-loss spectroscopy (EELS) and
high resolution transmission electron microscopy (HRTEM),
respectively, which result in large errors for the measured In
contents and to a lesser degree for the well widths. For instance,
when looking at the system with the lowest In content, grown at
$800^{\circ}$ C, the well width is $2.5\pm0.3$ nm and the In content
5$\pm3$\%. We therefore keep uncertainties in mind when comparing
with data from Ref.~\onlinecite{GrSo2005}.

For our atomistic theoretical framework, information on the In atom
distribution is also required. In the following we assume a random
distribution of In atoms, which is consistent with experimental
results, for instance, obtained by careful atom probe tomography
studies.~\cite{GrSo2005,MoPa2002,NeRo03,SmKa2003} We highlight that
even the 25\% In content sample is not expected to deviate from a
random In atom distribution as the atom probe analysis in
Ref.~\onlinecite{GaOl2008} shows. Also, the same assumption about
the random In atom distribution has been made by Watson-Parris
\emph{et al.}~\cite{Wats2011,WaGo2011} in the modified 3D
continuum-model. Since several studies have revealed that the alloy
microstructure significantly impacts the optical properties of
$c$-plane InGaN
QWs,~\cite{WaGo2011,ScCa2015,TaCa2016_RSC,AuPe2016,PiLi2017,JoTe2017}
we have constructed, for each In content, 175 different
configurations with different random In atom distributions. This
allows us to obtain reliable statistical averages and to calculate
quantities such as full width at half maximum (FWHM) of the PL
spectra and thus to compare this data with experiment.

Using Fermi's golden rule to calculate, as detailed above, the low
temperature and low carrier density optical spectra in combination
with the 175 different microscopic configurations, we find for all
but the 25\% In case smooth PL spectra; for the 25\% In system, we
still observe a slightly ``noisy'' PL spectrum. By using more
configurations, or a wider Lorentzian function for each PL peak
resulting from the different microscopic configurations, this
``noise'' could be mitigated; but, in order to keep the settings
consistent across the different In contents, such adjustments have
not been made. Thus, the predicted FWHM values for the 25\% In well
should be regarded as a lower bound for this system. Since we want
to study the electronic structure and the optical properties in
detail and for different temperatures, we consider for higher
temperatures and carrier densities not only ground states of each of
the microscopic configurations, but also excited states. Thus, we
have calculated for each of the 175 different configurations per In
content 20 electron and 40 hole states. With this we are able to
account for the fact that carriers will populate excited states at
elevated temperatures.

As a last ingredient we introduce structural inhomogeneities in the
theoretical description, namely well width fluctuations (WWFs) at
the upper interface of the well.~\cite{Wats2011,WaGo2011,ScCa2015}
Following the experimental and theoretical work in
Refs.~\onlinecite{Wats2011,WaGo2011,ScCa2015}, these WWFs are
described by disk-like objects with a diameter of approx. 5 nm and a
height of 2 monolayers. It should be stressed that these objects
allow now, in a spatially restricted region, the presence of In
atoms in the GaN barrier. Given that we treat InGaN as a random
alloy, the actual shape and form of these WWFs will vary and change
from microscopic configuration to microscopic configuration. It
should be noted that this situation is different when, for instance,
compared to a continuum-based calculation, where the WWF would be
rigid object with constant size and shape.

Effects such as penetration of In atoms into the barrier, beyond the
above discussed WWFs, have not been included. Neglecting these
effects is based on the following arguments given in the literature.
First, it has recently been shown that by a careful choice of the
growth conditions this effect can be reduced.~\cite{MaPi2017} While
this In penetration effect could be expected to be of relevance for
the samples of Ref.~\onlinecite{GrSo2005}, in structures grown more
recently this effect is expected to be less pronounced. Moreover,
Watson-Parris studied such a diffuse upper QW interface and
concluded that this factor has no noticeable impact on carrier
localization effects.~\cite{Wats2011} This stems from the fact that
the distribution of the In atoms, when penetrating into the barrier,
is homogeneous in the QW plane while in the observed WWFs the In
content is more noticeably concentrated. Thus, the impact of the
WWFs on carrier localization characteristics is much more pronounced
than a diffuse upper interface. Finally, for a consistent comparison
between the results of our model here and those of the modified 3D
single-band effective mass model,~\cite{Wats2011,WaGo2011} the same
basic assumptions about the structural properties have been made.
Following Ref.~\onlinecite{WaGo2011} and based on the other
considerations above, we exclude in our model In atom penetration
into the barrier, except for effects introduced by WWFs.

\section{Results}
\label{sec:Results}

Having introduced the theoretical framework and the InGaN/GaN QW
systems to which it is applied, we present in the following
subsections the results of our calculations. In a first step, in
Sec.~\ref{sec:General_aspects}, we give an overview of some general
aspects of the electronic structure of $c$-plane InGaN/GaN QWs.
Special attention is paid to carrier localization effects. This
analysis will support and underpin the findings of the following
sections. In Sec.~\ref{sec:TheoVsExp} we use the experimental data
extracted from Refs.~\onlinecite{GrSo2005} and~\onlinecite{Wats2011}
to benchmark our theoretical results across the In content
composition range from 5\% up to 25\% and compare also to literature
theory data by Watson-Parris~\emph{et al.}~\cite{Wats2011,WaGo2011}
(see above). In the following section,
Sec.~\ref{sec:CarrierLocLength}, we study localization lengths as a
function of the In content and temperature for a fixed carrier
density. Our calculated values are compared with experimental
literature data from the same study~\cite{GrSo2005} discussed in
Sec.~\ref{sec:TheoVsExp}. Finally, in Sec.~\ref{sec:WFoverlaps}, the
electron hole wave function overlaps are investigated. Attention is
directed towards the contribution of the in-plane carrier separation
to the reduction in the wave function overlap with increasing In
content. Both Secs.~\ref{sec:CarrierLocLength}
and~\ref{sec:WFoverlaps} will shed light on the connection
between carrier localization effects and \emph{changes} in the
optical properties of InGaN/GaN QWs when pushing the emission from
the blue wavelength regime to the green and orange part of the
spectrum.

\subsection{General aspects of the electronic structure of $c$-plane InGaN/GaN
quantum wells} \label{sec:General_aspects}

\begin{figure}[t!]
\includegraphics{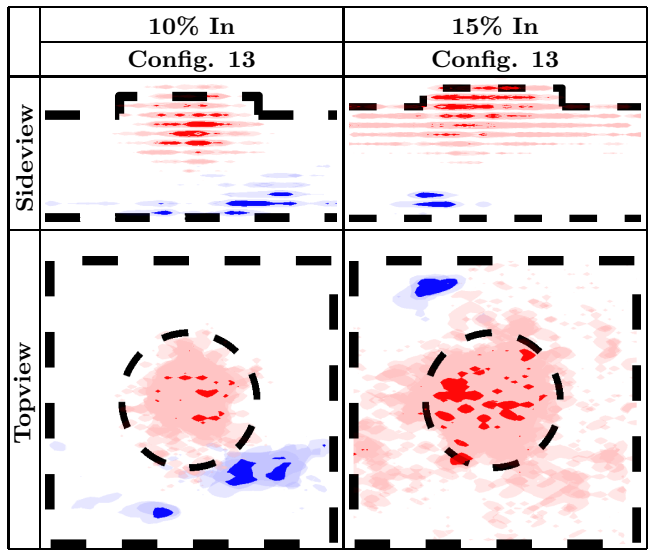}
\caption{Isosurface plots of electron (red) and hole (blue) ground
state charge densities. The light (dark) isosurfaces correspond to
15\% (40\%) of the respective maximum values. Results are given for
Config. 13 for $c$-plane InGaN/GaN QWs with 10\% (left) and 15\% In
(right) in the well. The dashed lines indicate the QW interfaces.
The results are shown for a sideview (perpendicular to $c$-axis) and
a topview (along the $c$-axis).} \label{fig:ChargeDens}
\end{figure}

Figure~\ref{fig:ChargeDens} shows isosurface plots of the electron
$|\psi^{e}_\text{GS}|^2$ (red) and hole $|\psi^{h}_\text{GS}|^2$
(blue) ground state charge densities for the microscopic
configuration (Config.) 13 of InGaN QWs with 10\% (left) and 15\% In
(right). The results are displayed for a sideview, perpendicular to
the wurtzite $c$-axis, and a topview, along the
wurtzite $c$-axis. Several aspects of these plots are 
of interest for the results discussed below. First, the side view
reveals a salient feature of $c$-plane InGaN QWs: the vertical
separation of the electron and hole wave functions due to the
built-in field. This significantly reduces the wave function
overlap. Second, for both systems we find strongly localized hole
ground state wave functions, revealing that random alloy
fluctuations are sufficient to give rise to carrier localization
effects. This localization of hole states is observed across all
configurations. For electrons we find that wave functions are
localized by the WWFs. In general, and also as in our previous
results,~\cite{ScCa2015,DaSc2016,TaMcM2018} the hole wave functions
are more strongly localized than those of the electrons. We will
return to this point later when discussing quantitatively the
electron and hole localization lengths.

Taking these two points together, our results show that in polar
InGaN/GaN QW systems the interplay between random alloy
fluctuations, WWFs, and electrostatic built-in field results in both
an out-of plane and an in-plane separation of the carriers. This
stems from the fact that holes do not necessarily localize below the
WWF where the electron is localized. We note that this in-plane
carrier separation is also present in structures without WWFs, as we
have discussed in previous studies, that the WWFs enhance, rather
than produce, this phenomenon.~\cite{TaMcM2018} The underling cause
of the in-plane separation is that, being separated to opposite ends
of the QW by the built-in field, the electrons and hole each inhabit
a different alloy/potential landscape, and those regions most
energetically favorable for an electron to localize at the upper
interface, are mainly uncorrelated with those regions energetically
favorable for the holes to localize at the lower interface (cf.
Fig.~\ref{fig:ChargeDens}). The finding of an in-plane spatial
separation is consistent with theoretical work of Watson-Parris
\emph{et al.},~\cite{WaGo2011} Auf der Maur \emph{et
al.},~\cite{AuPe2016} and Jones~\emph{et al.}~\cite{JoTe2017} It is
also consistent with the pseudo donor-acceptor pair model introduced
by Morel \emph{et al.}~\cite{MoLe2003} to explain time-resolved
optical properties of $c$-plane InGaN/GaN QWs. In
Refs.~\onlinecite{AuPe2016,Karp2017,Karp2018} by assuming a constant
defect related nonradiative contribution, the in-plane separation or
the combination with the built-in field is even regarded as the main
driver behind the green gap problem. However, before turning to this
question in more detail, we first start with benchmarking our theory
against experimental data across the In composition and the
resultant emission wavelength range relevant to the green gap
problem. We note that our low temperature experimental data
indicates (cf. Table~\ref{Tab:QW_results_EMA_EXP}) that 19\% In and
a well width of $3.2\pm0.2$ nm is enough to achieve emission in the
spectral range relevant to the green gap problem. In the theoretical
study of Ref.~\onlinecite{AuPe2016}, a well width of 3 nm and 30\%
In was required to push the emission wavelength to a comparable
wavelength, bearing in mind that these calculations were carried out
for emission at room temperature, the In content in the theoretical
study in Ref.~\onlinecite{AuPe2016} is noticeably different to the
In content in the experimental work in Ref.~\onlinecite{GrSo2005}
(cf. Table~\ref{Tab:QW_results_EMA_EXP}), while the well width is
not vastly different. Therefore, a detailed theory-experiment
comparison is important to ensure that the theoretical model
describes key optical properties of realistic and experimentally
relevant structures correctly before it is applied to analyze the
impact of carrier localization on the green gap problem.

\subsection{Theory vs. experiment: PL peak position energies and full
width at half maximum values} \label{sec:TheoVsExp}

The low temperature and low excitation carrier density optical
properties of InGaN/GaN QWs presented in Ref.~\onlinecite{GrSo2005}
provide an ideal starting point for comparing trends predicted by
our theory for PL peak energies and FWHM values over a wide In
composition range. In Table~\ref{Tab:QW_results_EMA_EXP}, the
optical and structural properties of the studied QW systems are
presented and summarized. As already discussed above, our aim is not
to match structural properties or PL peak energies exactly. Rather
we are interested in studying how the optical properties of polar
InGaN/GaN QWs \emph{change} with increasing In content and if we
capture the main features of the measured spectra. When comparing
the theoretical and experimental data presented in
Table~\ref{Tab:QW_results_EMA_EXP}, our theory gives a good
description of the PL peak energy with increasing In content.
Furthermore, we find good agreement with the theoretical data from
Refs.~\onlinecite{WaGo2011,Wats2011}, which used a modified 3D EMA
that also accounts for random alloy fluctuations.

In a second step, our calculated FWHM values are compared with both
the experimental~\cite{GrSo2005} and theoretical
data~\cite{Wats2011,WaGo2011} (cf.
Table~\ref{Tab:QW_results_EMA_EXP}). It is important to note that
the experimental optical studies have been carried out at low
temperatures $T$ ($T=6$ K) and low excitation carrier densities, so
that it is reasonable to assume that occupation of excited (electron
and hole) states is of secondary importance. Several features of
this comparison are now of interest. Looking at
Table~\ref{Tab:QW_results_EMA_EXP}, we observe that over the
composition range from 5\% to 15\% In (PL peak energies between
$\approx$2.6 eV and $\approx$3.35 eV), our theoretical results for
the FWHM are in good agreement with the experimental data from
Graham \emph{et al.}~\cite{GrSo2005} In this energy (wavelength
$\lambda$, In content) range an increase in the FWHM with decreasing
PL energy $E_\text{PL}$ (increasing emission wavelength $\lambda$)
is observed. The calculations from
Refs.~\onlinecite{Wats2011,WaGo2011} in this energy/composition
range give always lower numbers, except for the 5\% In case.
Nevertheless, over the 5\% to 15\% In content range the trend of
increasing FWHM values with decreasing PL peak energy (increasing In
content) is also borne out by the EMA. It should be noted that we
find good agreement between theory and experiment in terms of FWHM
values \emph{without} introducing In clustering effects, in contrast
to a recent study.~\cite{DiPe2019} As already indicated above, even
for the higher In content samples, we do not find experimental
evidence for clustering. Turning to the higher In content systems
(In content $>15$ \%) and lower PL peak energies, $E_\text{PL}<2.7$
eV, both theoretical models predict a further increase of the FWHM.
This is in contrast to the experimental data, which reveals a sudden
drop in the FWHM in the energy range below $E_\text{PL}<2.7$ eV
(samples with 19\% and 25\% In, cf.
Table~\ref{Tab:QW_results_EMA_EXP}). However, as we have highlighted
above, given that these single QW (SQW) systems have been
characterized by EELs and HRTEM, both nominal In content and well
width may exhibit large error bars. Turning to the literature,
recent results on multi QW (MQW) structures at room temperature show
that the FWHM increases at longer wavelength/lower PL peak
energies.~\cite{RoPr2018} Obviously, in these MQW systems other
factors such as well to well variations could play a role. Thus,
future studies, ideally with SQW samples of different In content,
could target further careful optical and structural characterization
of such systems, to shed more light on our observed discrepancies
between theory and experiment. However, such a detailed and refined
investigation is beyond the scope of the present study.

Overall, the comparison shows that our theoretical framework gives a
good description of several of the optical features of $c$-plane
InGaN/GaN QWs across the In content range relevant to the green gap
problem. Building on this, we now calculate carrier localization
lengths as a function of the In content. This quantity has recently
been used to establish a connection between (hole) localization
effects and the green gap.~\cite{Karp2017,Karp2018}

\subsection{Carrier localization length} \label{sec:CarrierLocLength}

Different approaches have been used in the literature to evaluate
the carrier localization length.  These include, amongst others: the
inverse participation ratio (IPR),~\cite{Th1974,Wegn80,LuRu2003} the
volume fraction,~\cite{ChLi2010} and the variance of the position
operator.~\cite{Wats2011} Here, we use the $\text{IPR}$ to calculate
wave function localization lengths. To define the IPR, we recall
that the TB wave function $\psi$ can be written as:
\begin{equation}
\psi=\frac{1}{\sqrt{N}}\sum^{N}_{i}\sum_{\alpha}a_{i,\alpha}\phi_{i,\alpha}\,\,
. \label{eq:TB_WF}
\end{equation}
The index $\alpha$ denotes the orbital type ($s$,$p_x$,$p_y$,$p_z$)
in our nearest neighbor $sp^3$ TB model and the label $i$ runs over
the $N$ lattice sites of the supercell ($N=81,920$). Based on
Eq.~(\ref{eq:TB_WF}), we can define the IPR as
follows:~\cite{TaCa2016_RSC}
\begin{equation}
\text{IPR}=\sum^{N}_i\left(\sum_\alpha|a_{i,\alpha}|^2\right)^2/\left(\sum^N_i\sum_\alpha|a_{i,\alpha}|^2\right)^2\,\,
. \label{eq:IPR_def}
\end{equation}
For a Bloch (delocalized) state the IPR would be
\mbox{$\text{IPR}=1/N$}.\cite{Wegn80} Conversely, a localized state
(only orbital contributions on the same lattice site $k$) would have
an IPR value of $\text{IPR}=1$. Given that a state with an IPR which
is twice that of another state can be regarded two times more
localized, we can use the volume of a reference state to associate a
given IPR with a length. In this we follow a procedure similar to
that of Thouless,~\cite{Th1974} who defined a localization length of
$l^\text{loc}=a\cdot\text{IPR}^{-1/3}$ for a cubic lattice with
lattice spacing $a$. However, we refine this procedure to account
for the fact that we are dealing with a wurtzite crystal structure
and more importantly with wurtzite QW systems which exhibit a
built-in field. Therefore, we differentiate between the in- and
out-of growth plane directions in these systems and go a step
further than the isotropic localization length defined in other
theoretical studies.~\cite{ChLi2010,Wats2011,Karp2017} This
treatment allows for a closer comparison between our theory and
experimental data that reports on in-plane hole localization
lengths.~\cite{GrSo2005}

To calculate the out-of plane localization length of a given state
$\psi$, we introduce the planar integrated probability density,
$P(z_n)$:
\begin{equation}
P(z_n) = \sum_{k,l}|\psi(x_k,y_l,z_n)|^2\,\,
, \label{eq:Planar}
\end{equation}
where $x_k$ and $y_l$ are the in-plane ($c$-plane) coordinates and
$z_n$ denotes the n$^{th}$ layer along the $c$-axis. This quantity
represents a probability density per layer, where $P(z_n)$ gives the
probability that the carrier described by state $\psi$ be found in
the n$^{th}$ layer of the supercell; $\sum_{n}^{N_{z}}P(z_n) = 1$,
where $N_{z}$ is the number of $x-y$-planes in the system. From this
one-dimensional probability density we may define an IPR value,
\mbox{$\textrm{IPR}_{z}= \sum_{z_{m}} (P(z_{m}))^{2}$} and follow
Thouless~\cite{Th1974} to relate this to a length via the formula:
$l^\text{loc}_{z}=a_{z}\cdot\text{IPR}_{z}^{-1}$, where $a_{z}$ is
the average spacing between $z$-layers in our relaxed supercells.

\begin{figure}[t!]
\includegraphics[width=\columnwidth]{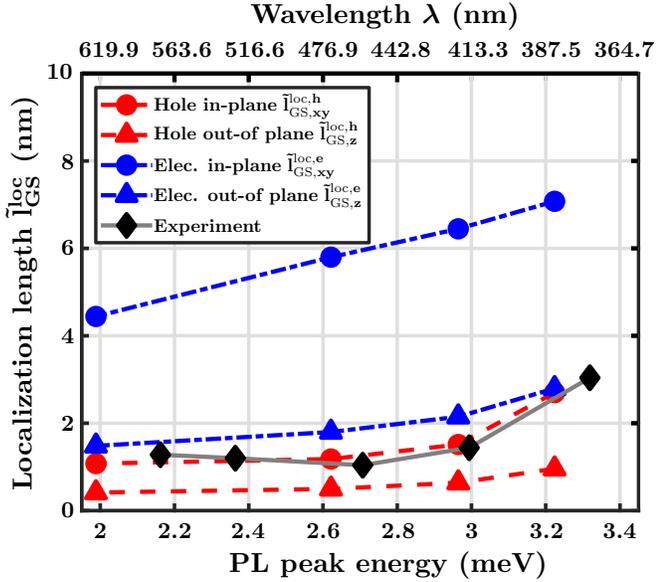}
\caption{Calculated average ground state electron
$l^{\text{loc},e}_\text{GS}$ (red symbols) and hole localization
$l^{\text{loc},h}_\text{GS}$ (blue symbols) displayed as a function
of PL peak energies. The in-plane (out-of plane) localization length
is given by the circles (triangles). The data is averaged over the
175 different microscopic configurations considered per In content.
Experimental data is given by the black diamonds.}
\label{fig:AVG_Loc_length}
\end{figure}

The in-plane localization length $l^\text{loc}_{xy}$ can be
calculated by building on the out-of plane length  and a reference
state. We do not go into the detail of the calculations here, but
present them in Appendix~A. Using this approach we have calculated
the in- and out-of plane \emph{average ground state} (GS)
localization lengths for electrons
$\tilde{l}^\text{loc,e}_\text{GS}(x)$ and holes
$\tilde{l}^\text{loc,h}_\text{GS}(x)$ by averaging over the 175
configurations for each In content $x$. The results are shown as a
function of the calculated PL peak energies in
Fig.~\ref{fig:AVG_Loc_length}. The top $x$-axis gives the
corresponding emission wavelength $\lambda$. Here and in subsequent
localization length plots, electron lengths are given in blue, and
holes in red; in-plane lengths are denoted by filled circles, and
out-of plane are denoted by filled triangles.

Figure~\ref{fig:AVG_Loc_length} reveals three properties which are
general across all quantities. First, the hole ground state wave
functions are more localized than those of the electrons, in both
the in- and out-of plane dimensions; this reflects the behavior seen
in Fig.~\ref{fig:ChargeDens} and owes to the larger effective mass
of the holes. Secondly, we note that there is a noticeable
difference between all in- and out-of plane localization lengths;
this highlights shortcomings which may be incurred by utilizing an
isotropic localization length, as previous studies
have.~\cite{Karp2018,ChLi2010} Thirdly, we highlight that all
localization lengths reduce with decreasing PL peak energy, but that
the rate of decrease is different for different quantities.

Turning to the specifics for the holes first, we note the in- and
out-of plane localization lengths change very little for emission
energies below 3 eV. The situation is slightly different for
electrons, at least for the in-plane localization length. Here with
increasing In content/wavelength the electron groundstate in-plane
localization length decreases more strongly. We impute this to the
strong impact that the combination of the built-in field and WWF has
on the ground state electron wave function: as the field increases,
the electron wave function is pushed more into the WWF and as a
consequence is more laterally confined. We note two aspects of the
QWs that affect this behavior: firstly, the WWFs and secondly the
increasing well width with increasing In content. Keeping the well
width fixed between the systems with 15\% In and 25\% In at, for
instance, $L_w=3$ nm would reduce the QCSE in the 25\% In case. In
addition if WWFs would then also be absent, the electron
localization length should increase compared to the data presented
here; this in general would come into play for the higher In content
regime where WWFs are more important. We note also that a very good
agreement between our calculated hole in-plane localization lengths
and data extracted experimentally from Huang-Rhys factor
measurements is achieved,~\cite{GrSo2005} adding further trust that
carrier localization effects are treated accurately in our modeling
frame. Overall, this emphasizes the importance of benchmarking the
theoretical model against experimental data so that carrier
localization effects are not over- or underestimated.

So far we have only calculated ground state localization length
values. Such an analysis should be sufficient when comparing our
results to low temperature and low carrier density (below 10$^{11}$
electron hole pairs cm$^{-2}$) experimental data.~\cite{ChSc2018}
With increasing temperatures and carrier densities, energetically
higher lying states become important. To shed light on this, we have
calculated the mean localization length
$\overline{l^\text{loc,$\lambda$}_\alpha}(x,T)$, with $\alpha=xy$
(in-plane) or $\alpha=z$ (out-of plane), for both electrons
($\lambda=e$) and holes ($\lambda=h$) for low temperatures $T$ (here
$T=10$ K) and at room temperature ($T=300$ K). All the following
calculations have been carried out for a fixed sheet carrier density
of $1.5\times10^{12}$cm$^{-2}$. We have chosen this carrier density
to go beyond the low carrier density assumption, but at the same
time to avoid entering the regime where emission from a higher
energy band is observed in experimental studies, which stems from
saturation of localized (ground) states.~\cite{ChSc2018} In future
studies, higher density regimes shall be targeted. However, to
analyze trends in the average localization length and to gain
insight into the impact of excited states on this quantity, our
present approach is a good starting point.

\begin{figure}[t!]
\includegraphics[width=\columnwidth]{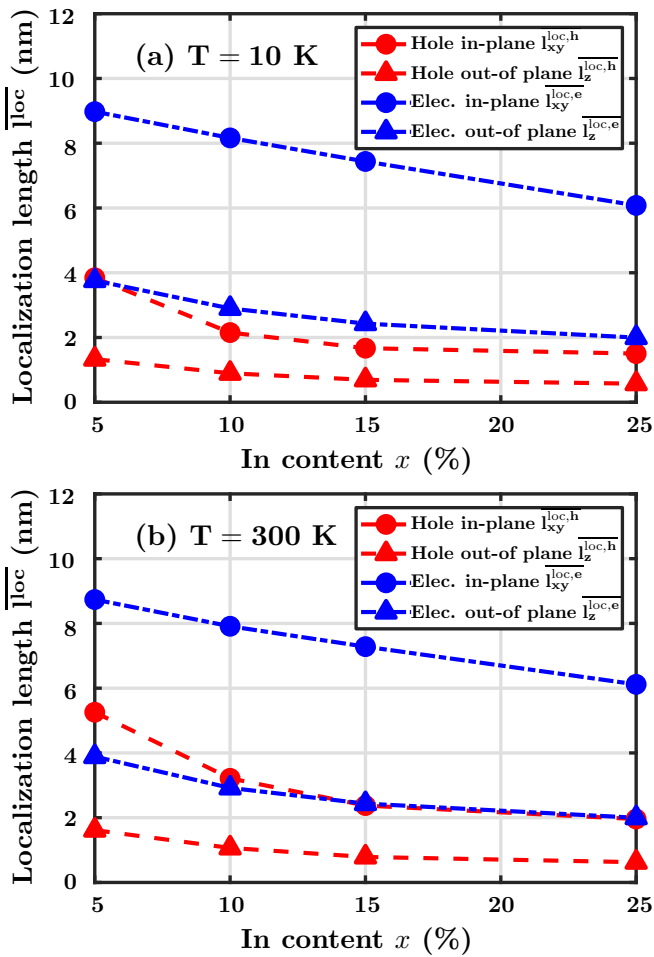}
\caption{In- (filled circles) and out-of plane (filled triangles)
electron (blue) and hole (red) localization lengths at (a) a
temperature of $T=10$ K and (b) $T=300$ K. The sheet carrier density
is $1.5\times10^{12}$ cm$^{-2}$. Calculations have been average over
the 175 configurations.} \label{fig:Loc_len_Carrier_Den}
\end{figure}

\begin{figure*}[t!]
\includegraphics{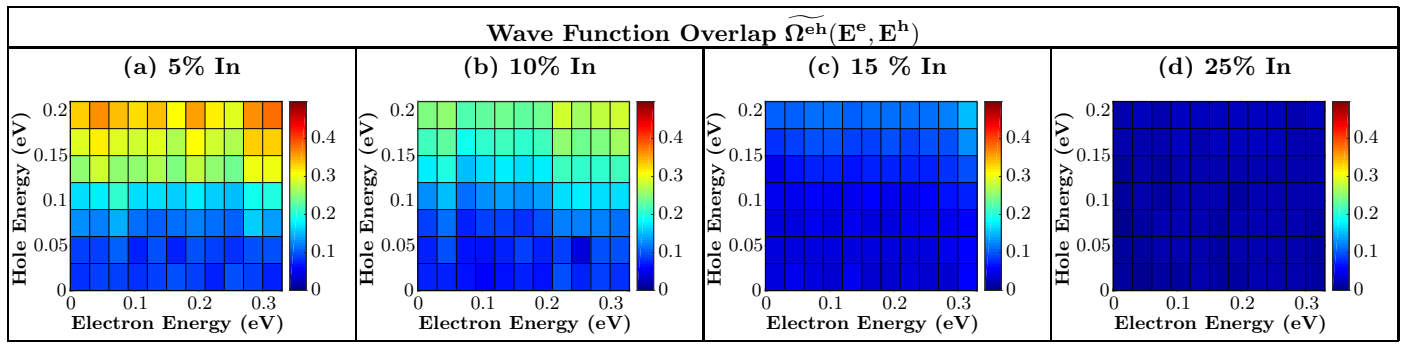}
\caption{Energy resolved average modulus electron-hole wave function
overlap $\widetilde{\Omega^{eh}}(E^{e},E^{h})$ for the considered
four different In contents, namely (a) 5\%, (b) 10\%, (c) 15\% and
(d) 25\%. The electron energies $E^{e}$ are given on the $x$-axis
while the hole energies $E^{h}$ are given on the $y$-axis. The
results have been obtained by averaging over 175 configurations per
In content, where an energy bin size of 30 meV has been used. The
data is plotted with respect to the corresponding conduction band
and valence ``band'' edges.} \label{fig:Overlap_EH_full}
\end{figure*}

Therefore, using our calculated energy distribution of electron and
hole states and Fermi-Dirac statistics,
$\overline{l^\text{loc,$\lambda$}_{\alpha}}(x,T)$ is calculated via:
\begin{equation}
\overline{l^\text{loc,$\lambda$}_{\alpha}}(x,T)=\frac{1}{N_\text{C}}\sum^{N_\text{C}}_{i=1}\sum^{N_\lambda}_{j=1}f^\lambda_{j}(E_{j,i},x,T)l^\text{loc,$\lambda$}_{\alpha,j,i}(E_{j,i},x)\,
, \label{eq:Loc_Temp}
\end{equation}
where $f^\lambda$ denotes the Fermi-Dirac distribution for electrons
($\lambda=e$) or holes ($\lambda=h$); $N_c$ is the total number of
microscopic configurations ($N_c=175$) and $N_\lambda$ denotes the
number of states ($N_e=20$; $N_h=40$). Given that our calculations
show carrier localization effects, and that the wave vector is in
general not a good quantum number, $f^\lambda$ is treated as in the
case of a quantum dot system with discrete energy levels and not a
$\mathbf{k}$-dependent energy dispersion.~\cite{WuLi2009} In doing
so, Eq.~(\ref{eq:Loc_Temp}) allows us to obtain mean localization
lengths at a given temperature $T$, carrier density and In content
$x$. Further information, including the evolution of the hole
localization length with energy and In content, is given in
Appendix~A.

Our results as a function of the In content $x$ in the well are
shown in Fig.~\ref{fig:Loc_len_Carrier_Den} (a) for low temperatures
($T=10$ K) and in (b) for room temperature ($T=300$ K). Firstly, we
see that carrier localization effects are still pronounced at higher
carrier densities and temperatures. Comparing
Fig.~\ref{fig:AVG_Loc_length} with
Fig.~\ref{fig:Loc_len_Carrier_Den} (a) we find that the increased
carrier density leads to an increase in all localization lengths. In
the case of the holes this is attributable to the population of
energetically higher-lying localized states (see also Appendix~A for
more details), whilst for the electrons the increase in length may
be due to occupation of states outside of the WWF, which exhibit
larger localization lengths due to the reduced (in-plane)
confinement. In the absence of WWF a slight increase in the electron
in-plane localization may be expected. Comparing
Fig.~\ref{fig:Loc_len_Carrier_Den} (a) with (b) we note that
increasing temperature has little effect on all the localization
lengths other than the in-plane hole localization (red filled
circles). Interestingly, we note that
$\overline{l_{xy}^{\text{loc},h}}$ for lower In contents are
increased more than $\overline{l_{xy}^{\text{loc},h}}$ at higher In
contents. For example, for the $T=300$ K case, on going from 5\% to
15\% In content, $\overline{l_{xy}^{\text{loc},h}}$ is reduced by a
factor of order 2.2, but on going from 15\% to 25\%, it is reduced
only by a factor of order 1.2. Thus the localization length change
in the regime relevant to the green gap problem shows negligible
variance when compared with lower In content regimes. We note that
the general trend of a smaller change in localization effects with
increasing In content has also been observed by other groups on
different quantities. For instance, Piccardo \emph{et
al.}~\cite{PiLi2017} observed that Urbach tail energies increase
strongly in the 6\% to 11\% In range, while for higher In contents
the Urbach tail energy changes very little.

We highlight that the saturation of the hole localization length in
the wavelength range relevant to the green gap (530 nm to 570
nm\cite{HaKa2015,AuPe2016}), is an important result given that this
quantity has generated interest as a first indicator of the
deleterious effects of alloy-induced localization effects for the
green gap problem. Given that there is no sudden \emph{change} in
the hole localization length with increasing In content/wavelenght
it is unlikely to explain the sudden decrease in IQE observed (cf.
Fig.~\ref{fig:MaxIQE_intro}) when going from blue to green emitters.

To extend this discussion, we note further important factors. We
recall that when growing InGaN/GaN QWs with higher In contents, the
growth temperature is lowered, resulting in an increase of the
(point) defect density.~\cite{GuZh2011,ArHe2012,UeIs2014} Point
defects have been demonstrated to act as nonradiative recombination
centers, which can lead to a reduction in the nonradiative
recombination lifetimes.~\cite{ChUe2005,LaPi2013,HaKa2015} Thus, the
holes, whose mean localization length remains small, may remain
isolated from these defects, except for those which are localized
near a defect. On the other hand electrons, whose (in-plane)
localization length is much larger, may now become increasingly
susceptible to an increased (point) defect density. As a consequence
nonradiative recombination rate may be increased. This contrasts
starkly with Refs.~\onlinecite{AuPe2016} and~\onlinecite{Karp2017},
where a constant, In content independent nonradiative recombination
rate is assumed, and the reduction in devices efficiency is mainly
attributed to built-in field and carrier localization effects due to
random alloy fluctuations.

Finally, we note that if the carrier densities are increased further
than those examined here, additional effects like field screening
will come into play, and the localization lengths of especially the
electrons will further increase, as we have shown in a recent
work.~\cite{TaMcM2018} The carriers will thus become more
susceptible to defects, and the aforementioned conclusions will be
strengthened. In the absence of WWFs the electron localization
length should increase and thus further supporting this conclusion.
An interesting aspect of the above interpretation is the significant
role played by the electrons, raising the question of whether it is
the electrons which are the important carriers for the explanation
of the green gap, rather than the holes, which have traditionally
been targeted.

Having discussed the electron and hole localization lengths, we turn
now to the study of the wave function overlaps. In particular, we
will focus on the importance of in-plane carrier separation for the
green gap problem.

\subsection{Electron-hole wave function overlap} \label{sec:WFoverlaps}

To analyze how the electron-hole wave function overlap changes with
In content, without using a VCA system as a reference, we proceed as
follows. In a first step, for each of the 175 different microscopic
configurations per In content, the \emph{modulus} electron hole wave
function overlap, given by
\begin{equation}
\Omega^{eh}(n,m)=\sum^{N}_{i}(|\psi^{e}_{i,n}|^2|\psi^{h}_{i,m}|^2)^{\frac{1}{2}}=\sum^{N}_{i}|\psi^{e}_{i,n}||\psi^{h}_{i,m}|\,\,
, \label{eq:Overlap_full}
\end{equation}
has been calculated between electron state $\psi^{e}_{n}$ and hole
state $\psi^{h}_{m}$. The sum runs over all the $N$ lattice sites of
the supercell. We note that our aim is to study the contribution of
the in-plane carrier separation to the reduction of the total wave
function overlap; radiative recombination lifetimes shall be
targeted in future studies. Furthermore, we stress that our goal is
to analyze the in-plane carrier separation \emph{without} using a
VCA as a reference, which is often done.~\cite{AuPe2016,JoTe2017}
While a comparison between an atomistic model and a VCA model would
allow us, in principle, to distinguish between in- and out-of plane
separation contributions (carrier localization and built-in field),
this comes at cost: the magnitude and relative importance of the two
different factors (built-in field and carrier localization), depend
on how the parameters of the VCA calculations are are determined
(linear interpolation vs. bowing parameters). Thus the absolute
numbers from a comparison with VCA as a reference have to be treated
with care. An approach that targets this question without a VCA will
eliminate assumptions about interpolating material parameters and
remove therefore an extra layer of uncertainty
(over/underestimation) for the overall importance of the in-plane
carrier separation for the electron-hole wave function overlap. For
these reasons the modulus wave function overlap,
Eq.~(\ref{eq:Overlap_full}), provides a coherent and consistent
frame that does not suffer from assumptions made in a VCA reference
frame. Also the above approach allows us in a simple manner to study
the impact of the In content $x$ on the spatial separation of wave
functions in In$_x$Ga$_{1-x}$N/GaN QWs, both in- and out-of growth
plane. In general, more details about this approach and how we have
previously used it to gain insight into experimental observations
such as a carrier mobility edge or the appearance of a high energy
emission band can be found in Refs.~\onlinecite{ChSc2018}
and~\onlinecite{BlSc2018}, respectively. Furthermore, similar
approaches have been used by other groups to study the optical
properties of different nitride-based
nanostructures.~\cite{MaHa2013,MaGe2015}

Given that each electron and hole wave function is connected to a
corresponding eigenenergy, one can also study an energy resolved
modulus wave function overlap $\Omega^{eh}(E^{e}_n,E^{h}_m)$ where
$E^{e}_n$ is the electron eigenenergy and $E^{h}_m$ is the hole
eigenenergy of states $n$ and $m$, respectively. Therefore,
$\Omega^{eh}(E^{e}_n,E^{h}_m)$ can be visualized as a function of
electron $E^{e}_{n}$ and hole $E^{h}_{m}$ energies. In a second
step, the overlaps $\Omega^{eh}(E^{e}_n,E^{h}_m)$ from the 175
different configurations have been grouped together in energy bins
of 30 meV width. An averaged energy resolved modulus wave function
overlap $\widetilde{\Omega^{eh}}(E^{e},E^{h})$ is obtained by
dividing the sum of the overlaps in an energy bin by the number of
elements therein. To better compare the results from the different
In contents, the data is always plotted with respect to
corresponding conduction and valence ``band'' edges.

The results of this analysis are displayed in
Fig.~\ref{fig:Overlap_EH_full}, with the In content increasing from
left to right. Three main results may be inferred from these
figures: (i) there is an overall decrease in electron and hole wave
function overlap with increasing In content (and will width); (ii)
for a given system, the energy range over which electrons and holes
have low overlaps with each other depends on the In content; (iii)
the overlap depends more weakly on the electron energy than it does
on the hole energy.

Result (i) reflects mainly an increasing strain induced built-in
field (which separates carriers perpendicular to the plane of the
QWs, cf. Fig.~\ref{fig:ChargeDens}) and alloy-fluctuation induced
carrier localization (which separates carriers in-plane, cf.
Fig.~\ref{fig:ChargeDens}) with In content, as discussed in
Sec.~\ref{sec:General_aspects}. In our subsequent analysis we will
disentangle the contribution of these two effects, as a function of
the In content, in order to investigate the relevance of the alloy
induced in-plane carrier separation to the green gap problem.

Point (ii) is a manifestation of an earlier reported
result~\cite{TaCa2016_RSC} that, not only does the strength of
localization increase with In content, but so too does the energy
range over which it persists (see also Appendix~A for further
discussion and Ref.~\cite{TaCa2016_RSC}). This may have interesting
consequences for the green gap, or even efficiency droop: the
carrier densities and temperatures required to saturate the
localized states may be higher for higher In content QWs. We will
investigate this further below in the context of the green gap
problem.

Aspect (iii) results from the fact that the electron localization
length is large compared to that of the holes and changes little
when energetically higher-lying states are populated, whilst the
hole localization length increases as states deeper in the valence
``band'' are populated. More generally speaking, we note that this
changing localization character and dependence of overlap on energy
of electrons and holes away from their respective conduction or
valence ``band'' edges is also reflected in experimental studies,
for instance, in the appearance of a high energy emission band with
much shorter radiative recombination times at low temperature
($T=10$ K).~\cite{ChSc2018}

While Fig.~\ref{fig:Overlap_EH_full} gives first information on the
evolution of the wave function overlap with In content, it does not
provide information about the density of states. Therefore, to gain
further insight into all three aspects discussed above, in the next
steps, we include the effects of temperature and carrier density on
the wave function overlaps, and investigate the specific
contribution of the alloy-induced in-plane separation of carriers to
the reduction of the electron-hole wave function overlap,
$\Omega^{eh}(E^{e}_n,E^{h}_m)$.

First, we introduce carrier population effects due to higher
temperatures and carrier densities as before using Fermi-Dirac
statistics. Similarly to localization length,
$\overline{\Omega^{eh}}(T)$ is defined via:
\begin{equation}
\overline{\Omega^{eh}}(T)=\frac{1}{N_\text{c}}\sum^{N_\text{c}}_{\alpha}\sum_{i,j}f^e(T,E^{e}_{i,\alpha})f^h(T,E^{e}_{i,\alpha})\Omega^{eh}(E^{e}_{i,\alpha},E^{h}_{j,\alpha})\,\,
. \label{eq:mean_Omega}
\end{equation}
Here, $\alpha$ denotes the configuration number of the
\mbox{$N_\text{c}=175$} different microscopic configurations per In
content. The electron and hole energies are denoted by
$E^{e}_{i,\alpha}$ and $E^{h}_{j,\alpha}$. The factors $f^e$ and
$f^h$ are the Fermi-Dirac functions for electrons and holes,
respectively. The sheet carrier density $n_s$ for all calculations
is again $n_s=1.5\times10^{12}$ cm$^{-2}$.

Next, we examine the contribution of the in-plane separation to the
reduction of $\overline{\Omega^{eh}}(T)$ at low (T=10 K) and high
(T=300 K) temperatures. For the reasons discussed above we study
this question without using a VCA calculation. To achieve this
unified description, we make use of the planar integrated
probability density $P^{\lambda}_j(z_o)$, Eq.~(\ref{eq:Planar}).
From this quantity we can determine the planar integrated (modulus)
overlap $\overline{\Omega^{eh}_{P}}(T)$. This quantity reflects what
the overlap between two states would be if they were not separated
from each other \emph{in} the growth plane. Further discussion of
this quantity and the details of its derivation are given in
Appendix~B. We may quantify the amount by which this is larger than
$\overline{\Omega^{eh}(E^{e}_n,E^{h}_m)}$, or, more meaningfully,
the amount by which $\overline{\Omega^{eh}(T)}$ is reduced by
in-plane carrier separation, using the following difference:
\begin{equation}
\delta\overline{\Omega_\text{IP}}(T)=\frac{\overline{\Omega^{eh}_{P}}(T)-\overline{\Omega^{eh}}(T)}{\overline{\Omega^{eh}_{P}}(T)}\,\,
. \label{eq:Omega_InPlane}
\end{equation}
Thus $\delta\overline{\Omega_\text{IP}}(T)$ represents the
fractional reduction in the modulus overlap between electron and
hole wave functions due to their in-plane separation, i.e. due to
alloy and WWF induced carrier localization effects.

\begin{figure}
\includegraphics[width=\columnwidth]{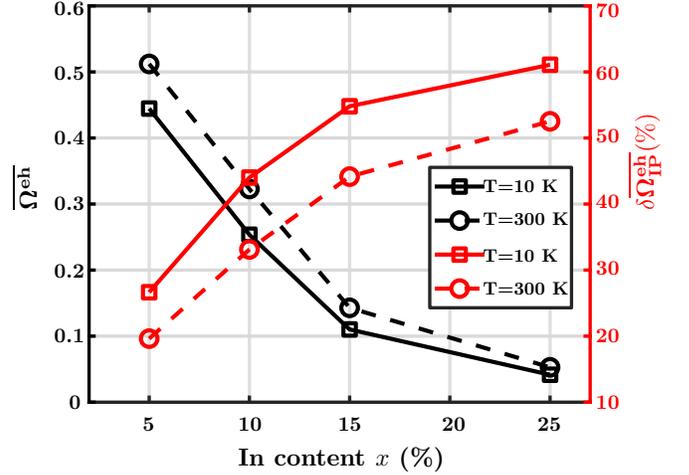}
\caption{The full mean modulus electron-hole wave function overlap
$\overline{\Omega^{eh}}$ is given in black; the in-plane
contribution to the spatial separation
$\delta\overline{\Omega^{eh}_\text{IP}}$ (in \%) is given in red.
Squares denote the results for a temperature of $T=10$ K, while the
circles indicate the data for $T=300$ K. The sheet carrier density
$n_s$ in the calculations is $n_s=1.5\times10^{12}\text{cm}^{-2}$.}
\label{fig:Overlap_Temp}
\end{figure}

Figure~\ref{fig:Overlap_Temp} depicts the results of this analysis
for the full overlap $\overline{\Omega^{eh}}(T)$ (black squares and
circles) and the reduction in overlap due to in-plane separation
$\delta\overline{\Omega^{eh}_\text{IP}}(T)$ (red squares and
circles). The squares denote $T=10$ K (low temperature) data, while
the results at $T=300$ K (room temperature) are given by the
circles.
Figure~\ref{fig:Overlap_Temp} reveals the general feature that with
increasing In content (and well width) $\overline{\Omega^{eh}}(T)$
decreases while $\delta\overline{\Omega^{eh}_\text{IP}}(T)$
increases. The latter indicates an increase in in-plane carrier
separation with increasing In content.

Turning specifically to the evolution of the mean \emph{full}
overlap $\overline{\Omega^{eh}}(T)$ (black circles and squares),
which gives a first indication of how the radiative recombination
rate changes with In content and temperature. Looking at the impact
of temperature on the results, we see that increasing the
temperature increases $\overline{\Omega^{eh}}(T)$  more for the
lower In content systems than it does for the higher In content
systems. This is a manifestation of this increasing energy range of
localized states with increasing In content, as confirmed by
Fig.~\ref{fig:Overlap_EH_full}. When increasing the temperature
energetically higher lying states are populated. On average and with
increased temperature, in the 5\% In cases, carriers start populate
more delocalized states sooner when compared to the 25\% case at the
same temperature and carrier density. Two important conclusions can
be drawn from this for the theoretical studies of the optical
properties of these structures. First, an accurate description of
the density of localized states is required for accurate modeling of
the radiative recombination properties. Second and connected to
this, the In content plays an important role. Very high In contents
may lead to an overestimation of carrier localization effects and
the energy range over which localization effects are relevant. This
highlights again the importance of our initial theory-experiment
benchmarking exercise.

Moreover, given that $\overline{\Omega^{eh}}(T)$ is related to the
radiative recombination rate, it can be seen that for instance for
the 15\% In case, the radiative recombination rate should increase
with increasing temperature. Such an effect has been observed in the
experimental study by Nippert \emph{et al.}~\cite{NiKa16} However,
previous theoretical studies failed to explain this
behavior~\cite{JoTe2017} or had to include In atom clustering
effects.~\cite{DiPe2019} Our results indicate the experimentally
observed trend with temperature, thus highlighting that the
distribution of localized states is well described by our
theoretical model.

Turning to the reduction in overlap due to in-plane carrier
separation, $\delta\overline{\Omega^{eh}_\text{IP}}(T)$, the data
clearly show that this factor increases with increasing In content.
While we find that indeed in-plane carrier separation contributes
strongly to the overall spatial separation of the electron and hole
wave functions and is at first glance especially important in the
blue to green (and even beyond) spectral range (15\% to 25\% In)
this contribution increases only very slightly
($\delta\overline{\Omega^{eh}_\text{IP}}(x=0.25,T)-\delta\overline{\Omega^{eh}_\text{IP}}(x=0.15,T)<$
10\%), when compared to the composition range 5\% In to 15\% In
($\delta\overline{\Omega^{eh}_\text{IP}}(x=0.15,T)-\delta\overline{\Omega^{eh}_\text{IP}}(x=0.05,T)\approx$
25\%). Thus, when considering \emph{changes} in the electron-hole
wave function overlap with increasing In content, rather than the
absolute numbers, the contribution for in-plane carrier separation
emitters operating in the blue to green spectral range is not very
different. This can be attributed to the situation that the
distribution of localized states with energy are not vastly
different over the ranges of the considered temperatures and carrier
density. A further discussion is given in Appendix~B. We point out
that such a small change in carrier localization effects in the
higher In content regime has also been observed by other groups on
other quantities such as Urbach tail energies.~\cite{PiLi2017}

Finally, we note again that the argument of a smaller change in
$\delta\overline{\Omega^{eh}_\text{IP}}$ for higher In contents,
when compared to the lower In content systems, would only be
strengthened when keeping the well width fixed for different In
contents and when neglecting WWFs. This again goes back to the
situation that especially for the higher In content systems the QCSE
is reduced when for instance assuming a constant well width of e.g.
3 nm. Thus for the higher In content system (25\% In) the overall
wave function overlap should increase, compared to the data
presented for this system. Secondly, neglecting the WWFs and
reducing the QCSE by reducing the well width of the here chosen 25\%
In structure, so that it is similar to the 15\% case, the in-plane
carrier localization of electrons will be reduced, which yields a
reduction of the in-plane carrier separation between electrons and
holes. This will lead to an overall reduction of
$\delta\overline{\Omega^{eh}_\text{IP}}$ and an even smaller change
in $\delta\overline{\Omega^{eh}_\text{IP}}$ in the 15\% to 25\% In
content range when removing WWFs and assuming the same well width
e.g. 3 nm.

Taking our results all together, it casts into doubt the idea that
the observed slight change in
$\delta\overline{\Omega^{eh}_\text{IP}}$ with In content in the
longer wavelength regime is the fundamental origin of the green gap
problem. In other words, if the in-plane carrier separation presents
a significant problem for the efficiency of InGaN-based light
emitters, one could expect the drop of in efficiency to be between
the near UV and the blue spectral range, with blue and green being
almost equally efficient/inefficient. However, this is not observed
in practice.

\begin{figure}
\includegraphics[width=\columnwidth]{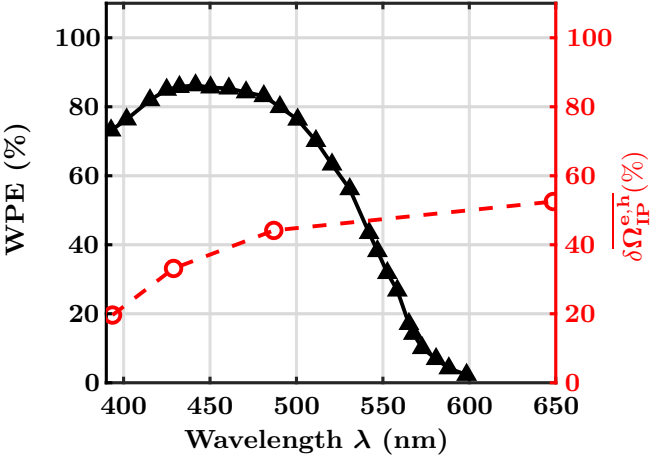}
\caption{Literature wall plug efficiency (WPE) values (black stars)
of InGaN QW systems as a function of the emission
wavelength.~\cite{WPE_2019} The calculated reduction in wave
function overlap due to in-plane carrier separation effects is given
by the open red circles.} \label{fig:MaxIQE}
\end{figure}

To further illustrate this point, we plot
$\delta\overline{\Omega^{eh}_\text{IP}}(T)$ against experimental
literature data on peak wall plug efficiency (WPE) values of
InGaN/GaN QW structures operating in the blue to green spectral
region. The experimental WPE values (black filled triangles) are
shown in Fig.~\ref{fig:MaxIQE} as a function of the emission
wavelength. The data has been extracted from data given in
Ref.~\onlinecite{WPE_2019}. The black line serve as a guide to the
eye. For comparison we have also plotted
$\delta\overline{\Omega^{eh}_\text{IP}}(T)$ at \mbox{$T=300$ K} as a
function of the calculated emission wavelength $\lambda$ of the
InGaN/GaN QWs with different In contents. The PL peak energies
originally calculated at low temperature ($T=0$ K) and given in
Table~\ref{Tab:QW_results_EMA_EXP} have been adjusted by using the
following correction~\cite{ElPe1997,Elis2003}
\begin{equation}
E_\text{PL}(T)=E_\text{PL}(0)-\frac{\alpha T^2}{\beta
+T}-\frac{\sigma^2}{k_bT^2}\,\, , \label{eq:Temp_correction}
\end{equation}
to account for the fact that experimental data in general is
obtained at room temperature ($T=300$ K). $E_\text{PL}(T)$ denotes
the energy of the PL peak energy at temperature $T$ and
$E_\text{PL}(0)$ is the PL peak energy at \mbox{$T=0$ K}. The second
term on the right-hand side corresponds to the temperature
dependence of the band gap described by the Varshni
equation,~\cite{VuMe2003} where $\alpha$ and $\beta$ are parameters
taken from Ref.~\onlinecite{VuMe2003}; a linear interpolation
between InN and GaN values has been used. The last term on the
right-hand side of Eq.~(\ref{eq:Temp_correction}) accounts for
thermal distribution of carriers between localized states. Here,
$k_b$ is the Boltzmann constant and $\sigma$ is connected to the
red-shift of the PL peak position with temperature (``S''-shape
temperature
dependence).~\cite{ElPe1997,Elis2003,HaWa2012,WaBa2001,ZhLu17} It
should be noted that the ``S''-shape temperature dependence of the
PL peak energies has two red-shift regions and that
Eq.~(\ref{eq:Temp_correction}) can only be used for the  high
temperature part i.e above the temperature of the maximum FWHM
value.~\cite{Elis2003} For $\sigma$ we have used values reported in
the literature for InGaN/GaN QWs with In contents ranging from 11\%
to 23\%.~\cite{HaWa2012,WaBa2001,ZhLu17} For the studied 5\% In
system we have used a $\sigma$ value of $\sigma=10$ meV. Using
Eq.~(\ref{eq:Temp_correction}) with literature $\sigma$-values
allows us to obtain a first approximation of the emission energies
at elevated temperatures $T$. Calculating $\sigma$ values from our
data or a full PL spectrum near room temperature is beyond the scope
of the present study.

Starting our analysis with the peak WPE values (black filled
triangles), we observe first that the WPE is approximately constant
in the $\lambda=420$ nm to $\lambda=490$ nm window. In contrast to
this, a strong decrease in these values from a wavelength of
approximately $\lambda=490$ nm to \mbox{$\lambda=580$ nm} is
observed. Comparing this now with our calculated in-plane separation
induced overlap reduction contribution,
$\delta\overline{\Omega^{eh}_\text{IP}}(T=300 \text{K})$, we find
that $\delta\overline{\Omega^{eh}_\text{IP}}(T)$ increases more
rapidly in the $\lambda=400$ nm to $\lambda=490$ nm regime. However,
this is the range in which the WPE is approximately constant. On the
other hand the opposite behavior for WPE and
$\delta\overline{\Omega^{eh}_\text{IP}}(T=300 \text{K})$ is observed
in the wavelength window $\lambda>490$ nm, meaning that
$\delta\overline{\Omega^{eh}_\text{IP}}(T=300 \text{K})$ shows only
a slow increase with wavelength while the WPE drops quickly.
This is the opposite to what would be expected if changes in the
in-plane carrier separation were the main driver behind the green
gap problem. Our results suggest instead that, given the slow rate
of change of $\delta\overline{\Omega^{eh}_\text{IP}}$, that carrier
localization, and the associated in-plane separation, contributes to
lesser extend to the drastic decrease in WPE. This indicates that
effects other than carrier localization must contribute to the sharp
reduction in WPE.

Furthermore, we note that if carrier localization effects were the
origin of the green gap problem, nonpolar InGaN/GaN QW systems
should \emph{not} exhibit this efficiency reduction for the
following reasons.~\cite{HuGr2017,MoRa2019} With the macroscopic
built-in field absent in these structures, the attractive Coulomb
interaction between electrons and holes leads to exciton
localization effects observed both in theory and
experiment.~\cite{GaSh2009,MaKe2013,ScTa2015,DaSc2016} In a
Hartree-type picture the in-plane separation between electrons and
holes will be strongly reduced due to their attractive Coulomb
interaction. However, as discussed by Monavarian \emph{et
al.}~\cite{MoRa2019}, nonpolar systems do not solve the green gap
problem. Obviously, nonpolar systems differ significantly in crystal
quality from their polar counterparts. This indicates already that
even when the in-plane carrier separation is strongly reduced in a
nonpolar system compared to a polar system, the crystal quality is
still very important.

While our calculations confirm that carrier localization effects and
thus the associated in-plane separation effects significantly
contribute to the wave function overlap reduction in $c$-plane
InGaN/GaN wells, we do not find strong \emph{changes} in this
quantity with changing In content in the In content range relevant
for the green gap problem. Thus it cannot explain the sharp decrease
in, for instance, WPE (cf. Fig.~\ref{fig:MaxIQE}) We highlight again
that previous theoretical studies, which assumed a constant defect
related nonradiative rate, attributed carrier localization to be the
main driver behind the green gap problem.~\cite{MaKa2017,Karp2018}
Our results favor other factors such as increased point defect
densities in the green spectral region and thus In content
\emph{dependent} nonradiative recombination pathways as the more
likely explanation for the green gap problem. A similar conclusion
has recently been drawn in Ref.~\onlinecite{DaYo2019}, concluding
that reducing the (point) defect density could close the green gap.

\section{Conclusion}
\label{sec:Conclusion}

In this work we have revisited the importance of carrier
localization for the green gap problem. While this topic has been
targeted by different groups, there remains uncertainty about the
primary source of this phenomenon, and thus the best route by which
it may be circumvented. As discussed above, several works impute the
green gap primarily to an increased defect
density,~\cite{LaKr11,LaHo13,DaYo2019} while other studies focus on
carrier localization effects as the main source of this
effect.~\cite{NiKa16,MaKa2017,Karp2018}

To shed new light onto this fundamental question we build our
analysis of carrier localization effects on an atomistic multi-band
tight-binding (TB) model that we rigourously benchmark against
experimental data for InGaN/GaN QWs emitting over a wide range of
wavelengths. These studies reveal that carrier localization effects
are accurately described by our model. Quantities such as in-plane
carrier separation and carrier localization lengths, originating
from random alloy fluctuations, for In contents between 5\% and 25\%
are directly calculated in an atomistic frame without fitting to
experimental data.

Equipped with this model we show that carrier localization effects
in InGaN QWs do not change strongly when the In content in the well
is varied between 15\% and 25\% and thus in the emission wavelength
range from to blue to green. This feature of a small change in the
localization effects with In content has not been considered or
discussed in works attributing the green gap problem to carrier
localization; only absolute numbers were discussed in previous work
rather than focussing on changes in the carrier localization
effects, while also keeping the defect related nonradiative
recombination rate constant with In content. Given that blue LEDs
are very efficient and that the differences in alloy induced carrier
localization effects do not change strongly between blue and green
emitters, only slight changes to the radiative recombination rate
are expected. Thus our findings cast doubt on the conclusion that
the nonradiative recombination due to defects is constant and In
content independent; given the small change of localization effects
in the content range relevant to the green gap problem, the sharp
decrease in wall plug efficiency would not be observed unless some
other quantity, like defect densities, increased sharply.

We come thus to the conclusion that further factors, such as the
connection between the large electron localization length and the
increased defect density in the high In content regime, which could
lead to reduced nonradiative carrier lifetimes, are more likely to
be the driver behind the green gap problem. Experimental studies
point in the same direction, although the details of the underlying
mechanisms could be somewhat
different.~\cite{LaKr11,HaKa2015,DaYo2019}

Based on this explanation, accounting for the importance of the
carrier (electron) localization length and increasing (point)
defects densities for the green gap problem, we suggest different
countermeasures. One such countermeasure is for instance the already
widely used approach of
underlayers.~\cite{ArBr2015,HaCa2017,ChKa2018} It has been shown
that via the use of underlayers the point defect density can be
reduced in the QWs, which should then be beneficial for the green
gap problem. Furthermore, on the finding of a relatively large
electron localization lengths and the potential connection to
nonradiative recombination rate, tailoring carrier localization
effects might be the way forward to close the green gap. Here, for
instance co-localization of carriers by introducing quantum dot like
structures in InGaN/GaN QWs could be a way forward. This would not
only have the benefit of reducing the electron localization length
but potentially also co-localizing electrons and holes in the same
spatial region. Experimentally this might be achieved by tailoring
the growth parameters (particularly the growth pressure) to induce
the formation of non-random clusters in QWs that are not usually
present (dot-in-well structures).

\begin{acknowledgements}
This  work  was  supported  by  Science  Foundation  Ireland,
Sustainable Energy Authority of Ireland (project numbers 17/CDA/4789
and 13/SIRG/2210) and the United Kingdom Engineering and Physical
Sciences Research Council (Grant No. EP/M010589/1).
\end{acknowledgements}

\section*{Appendix}

In this appendix we provide (i) additional information about the
calculation of the in-plane carrier localization length and its
evolution with energy and In content for holes and (ii) further
details of the in-plane carrier localization effects. In the
following section, the in-plane carrier localization length will be
discussed. Subsequently we discuss the in-plane carrier separation
in more detail.

\subsection{Hole localization length}
\label{sec:Details_Hole_loc}

In the main text we have discussed the calculation of the out-of
plane carrier localization length. Here we provide information about
the calculation of the in-plane localization length. To evaluate the
in-plane localization length $l^\text{loc}_{xy}$, we assign a single
value by use of a reference state and the underlying hexagonal
crystal symmetry. We choose as our reference state the electron
ground state of Config. 13 with 10\% In in the well, which is shown
in Fig.~\ref{fig:ChargeDens} (left). This choice is made on the
basis that in-plane the electron is mainly localized within the WWF.
We then associate this state with a cylindrical volume given by:
$V_\text{ref}= \pi\cdot r_\text{ref}^{2}\cdot h_\text{ref}$, where
$r_\text{ref}$ is the radius of the WWF ($r_\text{ref}=2.5$ nm) and
$h_\text{ref}=l^\text{loc}_{z,\text{ref}}$, the out-of plane
localization length for the reference state. Then, utilizing the
fact that the ratios of volumes occupied by two states are the same
as the inverse of the ratio of their IPRs, we obtain, for a given
microscopic configuration $n$, the following expression for the
in-plane localization length $l^\text{loc}_{xy}$ of a given state:
\begin{eqnarray}
 l^\text{loc}_{xy}
 &=& 2\cdot\sqrt{\left(\frac{\text{IPR}_\text{ref}}{\text{IPR}}\right)\left(\frac{r_\text{ref}^{2}\cdot
 l^\text{loc}_{z_\text{ref}}}{l^\text{loc}_{z}}\right)}\,\, .
 \label{eq:oof_length}
\end{eqnarray}
Here, $\text{IPR}_\text{ref}$ is the full IPR value defined by
Eq.~(\ref{eq:IPR_def}) for our reference state discussed above
(electron ground state of Config. 13; 10\% In). The factor of 2 in
Eq.~(\ref{eq:oof_length}) is to give the full length (diameter) of
the circular geometry assumed here.

In the main text of the manuscript we have focused on the mean
carrier localization length. In the following we present a detailed
study of the in-plane hole localization ($l^\text{loc,h}_{xy}$); the
out-of plane localization length for electrons and holes is mainly
dominated by the built-in field. Furthermore, we have seen in our
previous work that variations in the electron IPR values are much
smaller than those in the hole IPR values.~\cite{TaCa2016_RSC} In
what follows, we investigate the extent to which the hole in-plane
localization length changes as states deeper into the valence
``band'' are considered, since this can give insight/indaction of
the distribution of localized states in the valence ``band''
(density of states).

To address this point the in-plane localization length
$l^\text{loc,h}_{xy}$ has been calculated for each of the 40 hole
states in each of the 175 configurations per In content system. To
plot the data from different configurations as a function of energy,
the results have been collected in bins of size of 30 meV and the
average localization length per bin has been calculated. A bin size
of 30 meV was chosen so that it is not too large to mask any
intrinsic features but not so small as to produce any empty energy
bins. To compare the results between the different In contents
easily, the data is always shown with respect to the respective
valence ``band'' edge values.

\begin{figure}[t!]
\includegraphics[width=\columnwidth]{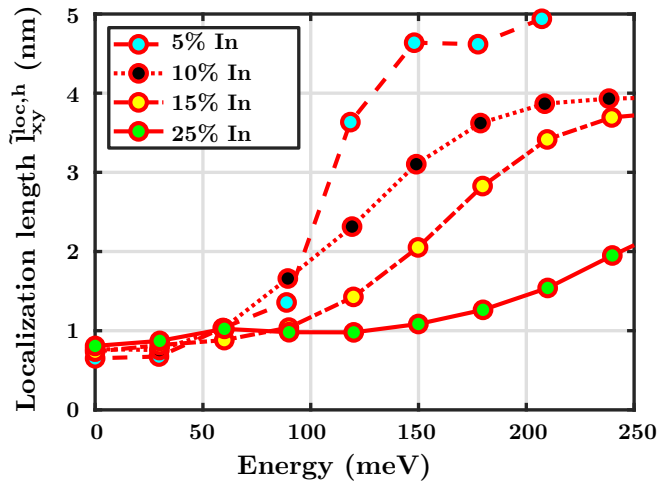}
\caption{Distribution of in-plane hole localization length
$l^\text{loc,h}_b$ with respect to valence ``band'' edge for
different In contents. The calculations use a bin size of 30 meV. 40
hole states per microscopic configuration have been calculated. The
binning has been done over the 175 configurations.}
\label{fig:Loc_Len_dist}
\end{figure}

The resulting energy resolved hole in-plane localization lengths are
displayed in Fig.~\ref{fig:Loc_Len_dist}. Independent of the In
content, we observe that the hole in-plane localization length
$l^\text{loc,h}_{xy}$ increases with increasing energy (moving
deeper into the valence ``band''). However, over an energy range of
approximately 60 meV there is not a huge difference in the hole
localization length at each In content. At first glance this seems
to be in contradiction with Fig.~\ref{fig:AVG_Loc_length} where the
average \emph{ground state} in-plane localization length
$\tilde{l}^\text{loc,h}_\text{GS,xy}$ is plotted and at 5\% In
($E^\text{theory}_\text{PL}$=3.23 eV)
$\tilde{l}^\text{loc,h}_\text{GS,xy}$ is larger compared to
$\tilde{l}^\text{loc,h}_\text{GS,xy}$ of the 25\% In case
($E^\text{theory}_\text{PL}$=1.99 eV). It is important to note that
in Fig.~\ref{fig:AVG_Loc_length} just \emph{averages over the ground
state values} are taken; this neglects any spread in the
corresponding hole ground state energies.
Figure~\ref{fig:Loc_Len_dist} shows an \emph{energy resolved}
in-plane localization length $l^\text{loc,h}_{xy}(E)$. Therefore,
one can find hole states that are as strongly localized in the 5\%
as in the 25\% In case over the same energy range from the
respective valence ``band'' edge. However, the density of these
strongly localized states increases significantly with increasing In
content. We have discussed this feature
previously,~\cite{TaCa2016_RSC} when studying IPR values of
electronic states as a function of the state number in $c$-plane
InGaN/GaN wells with varying In contents. Furthermore,
Fig.~\ref{fig:Loc_Len_dist} reveals that the energy region over
which strong localization exists increases with increasing In
content. For instance, while in the 5\% In case (blue filled
circles, dashed line) the in-plane localization length starts to
sharply increase at an energy of around 90 meV, for the 25\% In case
(green filled circles, solid line) we do not start to see any such
increase of the in-plane localization length until hole energies
exceed 210 meV.

\subsection{Wave function overlap and in-plane carrier localization}
\label{sec:Details_Wave_Overlap}

\begin{figure*}
\includegraphics{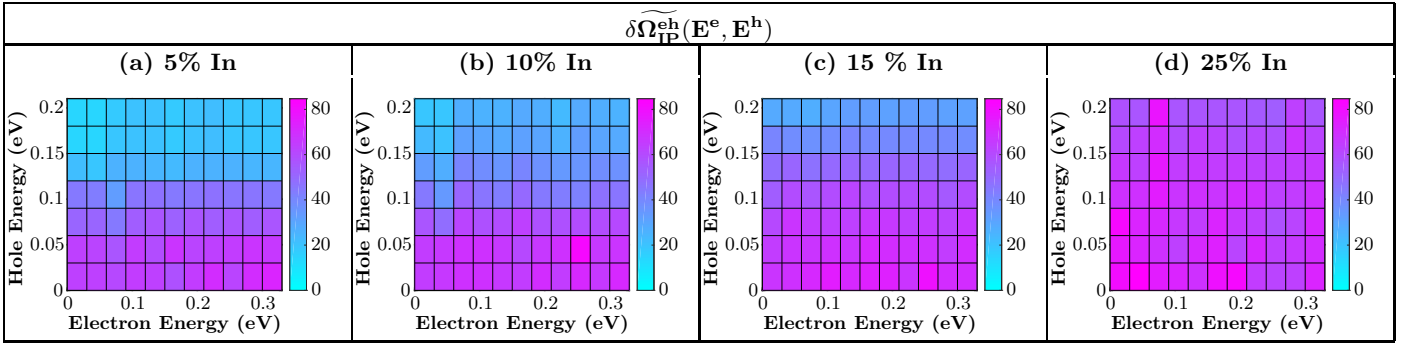}
\caption{Contribution $\delta\widetilde{\Omega^{eh}_\text{IP}}$ (in
percent) of the in-plane electron-hole separation to the electron
and hole wave function separation in the considered InGaN/GaN QWs
with In contents ranging from (a) 5\% up to (d) 25\%. Electron
energies $E^{e}$ are given on the $x$-axis and hole energies $E^{h}$
on the $y$-axis. Here, energies are plotted with respect to
conduction and valence ``band'' edges. The results have been
obtained by averaging over the 175 configurations per In content,
where an energy bin size of 30 meV has been chosen for both
electrons and holes.} \label{fig:Inplane_Overlap}
\end{figure*}

In the main text of the manuscript we have presented the
contribution of in-plane separation to the reduction in the wave
function overlap. Here, we provide the derivation of this quantity,
and detailed results underlying above presented mean values.

To obtain $\delta\overline{\Omega^{eh}_\text{IP}}$, we begin with
the planar integrated probability density $P^{\lambda}_j(z_o)$ of
Eq.~(\ref{eq:Planar}), which describes the one dimensional
probability density along the $c$-axis. We can use
$P^{\lambda}_j(z_o)$ to calculate the planar integrated (modulus)
overlap $\Omega^{eh}_P(n,m)$ via:
\begin{equation}
\Omega^{eh}_P(n,m)=\sum_{o}(P^{e}_n(z_o)\cdot
P^{h}_m(z_o))^{\frac{1}{2}}\,\, . \label{eq:Planar_eh}
\end{equation}

This quantity is sensitive only to the extent to which an electron
and hole wave function reside on the same plane, irrespective of
whether or not they occupy the same position in that plane. This is
because, based on Eq.~(\ref{eq:Planar_eh}), the in-plane extent of
the charge density is summed up into one value per plane.
Consequently, all calculations are carried out in the same frame
containing the same microstructure, local strain and built-in field
fluctuations as the full overlap calculations discussed above. The
calculated planar integrated overlaps $\Omega^{eh}_P(n,m)$ can again
be expressed in terms of electron $E^{e}_n$ and hole energies
$E^{h}_m$ and an averaged value, denoted by
$\widetilde{\Omega^{eh}_P}(E^{e},E^{h})$, which is obtained in the
same manner as before for $\widetilde{\Omega^{eh}}(E^{e},E^{h})$ and
discussed in the main text of the manuscript.

To quantify the amount by which this is larger than
$\Omega^{eh}(E^{e}_n,E^{h}_m)$, or, more meaningfully, the amount by
which $\Omega^{eh}(E^{e}_n,E^{h}_m)$ is reduced by in-plane carrier
separation, we utilize the following metric:
\begin{equation}
\delta\widetilde{\Omega_\text{IP}}(E^{e},E^{h})=\frac{\widetilde{\Omega^{eh}_{P}}(E^e,E^{h})-\widetilde{\Omega^{eh}}(E^{e},E^{h})}{\widetilde{\Omega^{eh}_{P}}(E^{e},E^{h})}\,\,
. \label{eq:Omega_InPlane}
\end{equation}

Figure~\ref{fig:Inplane_Overlap} displays
$\delta\widetilde{\Omega_\text{IP}}(E^{e},E^{h})$, in percent, as a
function of the electron and hole energies with respect to the
respective conduction and valence ``band'' edges.

It is apparent from the figure that in-plane separation strongly
impacts the electron-hole overlaps, with percentage reductions in
the overlap of up to around 80\% seen for all In contents.
Consistent with our results presented above on the localization
lengths, the magnitude and energy range of this reduction increases
with In content. For each In content,
$\delta\widetilde{\Omega_\text{IP}}(E^{e},E^{h})$ decreases  with
distance from the valence and conduction ``band'' edges, and
decreases faster for lower In contents than for higher. Again, the
delocalized and unchanging nature of the electron states are
evidenced by the relative insensitivity of
$\delta\widetilde{\Omega_\text{IP}}(E^{e},E^{h})$ to the electron
energy $E^{e}$, when compared to the hole energy $E^{h}$.

This significant impact of in-plane separation on the electron-hole
overlap is consistent with previous works which have highlighted the
negative impact of carrier localization and in-plane
separation.\cite{AuPe2016,Karp2017,Karp2018} However, as pointed out
earlier, it is important to study \emph{changes} in this quantity
between blue and green emitting systems if we wish to determine its
relevance to the green gap problem. If carrier localization and the
associated in-plane separation were the main driver behind the green
gap problem, then a strong increase in
$\delta\widetilde{\Omega_\text{IP}}(E^{e},E^{h})$ should be evident
when going from the 15\% to the 25\% In system.

Comparing the 15\% and 25\% In systems, we note that differences are
indeed apparent on the high-energy side of
Fig.~\ref{fig:Inplane_Overlap} (c) and (d); though, on the low
energy side, these differences are less marked. The question on the
importance of these differences in these energy regimes have been
targeted in the main text of the manuscript.

Finally, to obtain $\delta\overline{\Omega_\text{IP}}$(T) we combine
Eq.~(\ref{eq:Planar_eh}) and~(\ref{eq:mean_Omega}); the mean planar
integrated overlap $\overline{\Omega^{eh}_{P}}(T)$ is obtained
similarly; via Eq.~(\ref{eq:Omega_InPlane}) the reduction in
$\overline{\Omega^{eh}}(T)$ due to alloy induced in-plane
separation, $\delta\overline{\Omega_\text{IP}^{e,h}}(T)$, at a given
carrier density and as a function of temperature, is calculated.

\bibliography{../../../phdstef}

\end{document}